# Figure Rotation And The Mass of the Galactic Bulge

R. D. Blum

Department of Astronomy, The Ohio State University, 174 W. 18th Ave., Columbus, Oh, 43210

## ABSTRACT

The mass of the Galactic bulge, $M_B$, is estimated from the tensor virial theorem. By including the effects of a barred stellar distribution and figure rotation (81 km s$^{-1}$kpc$^{-1}$) and by assuming that the bar is oriented at $\theta = 20°$ to our line of sight, $M_B$ is found to be as high as $2.8 \times 10^{10}$ M$_\odot$. This estimate is in good agreement with the total mass inferred from the observed optical depth to microlensing for stars toward the inner Galaxy. For larger angles, $\theta \sim 40°$, or smaller pattern speeds (20 km s$^{-1}$kpc$^{-1}$), $M_B$ is found to be $\sim 1.7 - 1.9 \times 10^{10}$ M$_\odot$, similar to previous estimates for a barred bulge ($2 \times 10^{10}$ M$_\odot$) and axisymmetric bulge ($1.8 \times 10^{10}$ M$_\odot$).

*Subject headings:* Galaxy: center — Galaxy: kinematics and dynamics — Galaxy: fundamental parameters — Galaxy: structure

## 1. INTRODUCTION

Until recently, the axisymmetric mass model of Kent (1992) was viewed as the standard Galactic bulge model; it successfully fits observed kinematics and the near infrared surface brightness distribution over the inner few kpc of the bulge. However, analysis of existing and new photometric observations show that the bulge may be strongly triaxial. Blitz & Spergel (1991) found that the near infrared surface brightness distribution is asymmetric about $l = 0°$. Data from the DIRBE experiment on board the Cosmic Background Explorer (COBE) confirmed this result (Weiland et al. 1994). The light asymmetry has been interpreted as an asymmetry in the stellar distribution. Dwek et al. (1994) fit triaxial models to the light distribution and find that the bulge is strongly barred and has its long axis oriented at an angle $\theta$ of $\sim 10°$ to $40°$ to our line of sight. Dynamical modeling of observed gas kinematics in the inner Galaxy (Binney et al. 1991) also points to a barred mass distribution with $\theta \sim 16°$.

In addition to its inability to account for the observed light asymmetry, the Kent (1992) axisymmetric model appears to significantly underestimate the optical depth to microlensing for stars observed toward the inner Galaxy (Paczyński et al. 1994; Han & Gould 1994a; Zhao, Spergel, & Rich 1995). The Kent (1992) model leads to a prediction for the optical depth to microlensing which is a factor of $3-4$ lower than the observed optical depth reported by the Optical Gravitational Lensing Experiment (OGLE) (Udalski et al. 1994) and the Massive Compact Halo Object (MACHO) experiment (Bennett et al. 1994). A barred geometry increases the bulge's optical depth to self-lensing because the optical depth scales as the distance from source to lens. A barred geometry naturally places large numbers of stars at the near end of the bar (lenses) in front of stars at the far end of the bar (lensed sources). Accounting for a barred bulge increases the predicted optical depth by a factor of nearly 1.5 compared to the value for the Kent (1992) axisymmetric model (Han & Gould 1994a).

Barred models of the bulge produce predictions for $M_B$ which bracket the estimate for an axisymmetric bulge. Zhao (1994) and Han & Gould (1994a) estimate $M_B$ to be $2 \times 10^{10}$ $M_\odot$ and $1.6 \times 10^{10}$ $M_\odot$, respectively, for barred models of the bulge. This may be compared to $M_B = 1.8 \times 10^{10}$ $M_\odot$ for the Kent (1992) model (Kent does not state a value for $M_B$, but using his density law and derived mass to light ratio leads directly to this value). In this *Letter*, I demonstrate that a triaxial model of Dwek et al. (1994), but modified to account for the excess luminosity in the central 3°, and a predicted value of the bar pattern speed (Binney et al. 1991) lead to a bulge mass which may be a factor of 1.4 higher than that of Zhao (1994). In what follows, I adopt a value for the distance between the sun and Galactic center ($R_\circ$) of 8 kpc.

## 2. Mass Estimate

Given a density model and observational estimates of the kinetic energy and pattern speed of a rotating, barred stellar system, an estimate of the system's mass can be made from the tensor virial theorem. In the case of the Galactic bulge, sufficient kinematic observations exist to estimate the global kinetic energy of the system (see below), recent density models have been computed from the observed light distribution (Dwek et al. 1994), and an estimate of the pattern speed of the bulge potential has been made from observations of gas kinematics (Binney et al. 1991).

For a steady state, rotating, triaxial system, the tensor virial theorem results in the following balance between the system's kinetic and potential energy (Binney 1982):

$$-W_{xx} = 2T_{xx} + \Pi_{xx} + \omega^2(I_x - I_y)$$

$$-W_{yy} = 2T_{yy} + \Pi_{yy} - \omega^2(I_x - I_y) \qquad (1)$$

$$-W_{zz} = \Pi_{zz}$$

where $W_{ii}, T_{ii}$, and $\Pi_{ii}$ ($i = x, y, z$) are the components of the potential energy tensor, ordered motion kinetic energy tensor, and random motion kinetic energy tensor, respectively. These equations are for an inertial reference frame with the major axis along the $x$ axis and figure rotation with pattern speed $\omega$ about the $z$ axis. The $\omega^2$ term arises because the moments of inertia, $I_x$ and $I_y$ are not constant in time with respect to the inertial frame. This rotation adds an effective kinetic energy along the major axis, and takes an equal amount away from the intermediate axis. I have assumed there is no ordered motion in the $z$ direction.

I define the density distribution as $\rho = M_B \xi(x, y, z)$, where $M_B$ is the system mass and $\xi$ is the normalized density distribution. I ignore the disk mass; the effect of this is discussed in the next section. Next, I use the following results from Binney & Tremaine (1987):

$$I_x = \int \rho x^2 dV$$

$$W_{xx} = -\frac{1}{2}G \int \int \frac{\rho(\mathbf{r})\rho(\mathbf{r}')(x-x')^2}{|\mathbf{r}'-\mathbf{r}|^3} dV'dV$$

where $V$ is the volume, $\mathbf{r}$ and $\mathbf{r}'$ are spatial position vectors, and analogous equations hold for $y$ and $z$. Equation (1) may then be written as,

$$-\frac{1}{2}GM_B^2 w_{xx} = M_B(\overline{\sigma}_{xx}^2 + \overline{V}_x^2) + M_B\omega^2(i_x - i_y)$$

$$-\frac{1}{2}GM_B^2 w_{yy} = M_B(\overline{\sigma}_{yy}^2 + \overline{V}_y^2) - M_B\omega^2(i_x - i_y) \quad (2)$$

$$-\frac{1}{2}GM_B^2 w_{zz} = M_B\overline{\sigma}_{zz}^2$$

where $w_{ii} = W_{ii}/(0.5GM_B^2)$ and $i_i = I_i/M_B$. Note that the $w_{ii}$ and $i_i$ depend only on the adopted $\xi(x,y,z)$. The kinetic energy terms have been replaced by terms involving the system mass and mass weighted mean and random velocities ($\Pi_{ii} = \int \rho \sigma_{ii}^2 dV = M_B \overline{\sigma}_{ii}^2$; $2T_{ii} = \int \rho \overline{v_{ii}}^2 dV = M_B \overline{V}_i^2$).

To estimate the total mass, I adopt one of the best fitting triaxial distributions to the near infrared surface brightness distribution of the bulge from Dwek et al. (1994). This is their G2 (Gaussian) model which has principal axis scale lengths of 1487, 584, and 405 pc (for $R_o = 8$ kpc). This model is modified inside a major axis radius of 680 pc (415 pc minor axis radius) to include an axisymmetric density distribution as defined by Kent (1992). This hybrid model, similar to one employed by Zhao (1994), accounts for the higher observed luminosity in the central bulge which the G2 model does not. The two density laws were not smoothly joined, but agree reasonably well at the 680 pc radius. Numerical integration yields the values of the $w_{ii}$ and $i_i$.

The pattern speed has been estimated by Binney et al. (1991) to be 81 km s$^{-1}$ kpc$^{-1}$ from observations of the inner Galaxy gas kinematics.

The bulge mass is calculated from equation (2) after rotating from the $x,y,z$ frame to our line of sight. The angle between the bar major axis and our line of sight is taken as $\theta = 20°$, a value consistent with the G2 model of Dwek et al. (1994). Other models which also fit the light distribution well have inclinations up to $\theta \sim 40°$ (Dwek et al. 1994). Binney et al. (1991) derive a best fit value of 16°. After rotation, all that remains is to adopt a suitable value for the line of sight kinetic energy. Observations at angular distances from the Galactic center of 2°–7° show a large velocity dispersion between about 90 km

s$^{-1}$ and 130 km s$^{-1}$ (Rich 1990; Sharples, Walker, & Cropper 1990; Terndrup, Frogel, & Wells 1994; Blum et al. 1994). The mean velocities are generally observed to be small in comparison, except at larger longitudes (8°−10°) where they rise (Minniti et al. 1992; Harding & Morrison 1993) to 50−80 km s$^{-1}$. This increase is compensated for by a decrease in dispersion; the total kinetic energy at larger radii appears to remain at ∼ (100 km s$^{-1}$)$^2$. I adopt an average of the observed dispersion, (110 km s$^{-1}$)$^2$, for the line of sight kinetic energy.

The above assumptions lead to a total bulge mass of $M_B \approx 2.8 \times 10^{10}$ M$_\odot$. The estimate is most sensitive to the adopted angle ($\theta$) between the line of sight and bar major axis and the pattern speed. The $\omega^2$ term is proportional to $\cos(2\theta)$ when rotated from the $x, y, z$ frame to the line of sight. If $\theta$ is taken as 40°, the mass estimate falls to $1.9 \times 10^{10}$ M$_\odot$. If the pattern speed is 20 km s$^{-1}$ (and $\theta = 20°$), $M_B = 1.7 \times 10^{10}$ M$_\odot$.

## 3. DISCUSSION

The derived bulge mass is about 1.4 times larger than previous estimates based on a variety of techniques. The Kent (1992) dynamical model, which fits nearly all of the existing kinematic observations, results in a $1.8 \times 10^{10}$ M$_\odot$ bulge. The non-axisymmetry of the bulge coupled with rapid figure rotation allows for a much increased mass. I have ignored the disk contribution, which might result in an overestimate of the bulge mass. Applying the tensor virial theorem to the Kent (1992) density distribution and using the same observed kinetic energy estimate as for the bar model above, I find an axisymmetric bulge mass of $2.3 \times 10^{10}$ M$_\odot$. This suggests that ignoring the disk contribution can lead to a somewhat higher mass estimate, 20% in the axisymmetric case. The bar model is more compact than the Kent model, so the disk contribution may be less in that case. It is also possible that the adopted line of sight kinetic energy is too high. This is less likely since the observed kinetic energy appears to be $\gtrsim$(100 km s$^{-1}$)$^2$ out to ∼ 1500 pc (major axis radius), as discussed in the previous section.

Dwek et al. (1994) derive a stellar mass to light ratio for their bar model of the bulge using assumptions about the initial mass function and evolutionary history for bulge stars. Combined with their (luminosity) density distribution this gives a bulge mass of $1.15 \times 10^{10}$ M$_\odot$ (R$_\circ$ = 8 kpc). Zhao (1994) constructed a self consistent dynamical model based on the same Dwek et al. (1994) triaxial distribution used in this paper (and similarly modified in the central region). Zhao's dynamical model also fits a number of kinematic observations for a total bulge mass of $2 \times 10^{10}$ M$_\odot$. Part of the difference between Zhao's model and the virial estimate presented in this work is likely due to the fact that Zhao used a lower value of $\omega$ (60 km s$^{-1}$kpc$^{-1}$) which significantly reduces the $\omega^2$ term in equation 2. Han & Gould (1994a) also used the virial theorem to estimate the bulge mass (again, using the same modified Dwek et al. 1994 density distribution). However, they obtained a lower value (1.6×10$^{10}$ M$_\odot$) because they did not include the effect of figure rotation ($\omega = 0$ in their model). The $\omega^2$ term in equation 2 is approximately as large as the kinetic energy term if the pattern speed is 81 km s$^{-1}$kpc$^{-1}$. Wada et al. (1994) have recently modeled the inner Galaxy with a bar potential and run SPH simulations of the gas behavior under this potential. They derive a significantly lower pattern speed ($\sim$ 20 km s$^{-1}$) than Binney et al. (1991). In this case, the present mass estimate would be reduced to 1.7×10$^{10}$ M$_\odot$, a value comparable to that found by Han & Gould (1994a) and slightly less than that found by Zhao et al. (1995).

Since the predicted optical depth to microlensing is proportional to the total bulge mass (see, for example, Han & Gould 1994a), the mass estimate derived in section 2 would result in a predicted optical depth consistent (within the uncertainties) with the observed optical depth reported by OGLE (Udalski 1994) and MACHO (Bennett et al. 1994). For example, Han & Gould (1994a) report a predicted optical depth of $\tau = 0.95 \times 10^{-6}$ for a barred bulge. Increasing this value by the ratio of bulge masses (1.75) and adding the amount due to disk lensing ($\sim 0.5 \times 10^{-6}$, Paczyński 1991, Griest et al. 1991) results in $\tau = 2.2 \times 10^{-6}$. This value may be compared to the observed optical depth, $\tau \approx 3 \times 10^{-6}$, reported by Udalski et al. (1994) and Bennett et al. (1994). The observed optical depth may be uncertain by as much as 24 % (Han & Gould 1994b), so the prediction from the bar model is consistent with the observations.

Finally, note that the mass estimate made from the observed line of sight kinematics may be used to estimate the transverse kinematics. These estimates may be compared to existing observations. The longitudinal kinetic energy is predicted to be (153 km s$^{-1}$)$^2$. The proper motion study of Spaenhauer, Jones, & Whitford (1992) results in a longitudinal velocity dispersion of 119 km s$^{-1}$ ± 4 km s$^{-1}$ (R$_o$ = 8 kpc) for stars observed toward Baade's window ($l,b = 1°, -4°$). The predicted value appears rather high even though the observed value does not include the effects of mean motion (there was no proper motion standard to calibrate the zero point of the measurements). The latitudinal dispersion from Spaenhauer et al. (1992), which likely has a small mean motion, is 104 km s$^{-1}$ ± 4 km s$^{-1}$. This may be compared to the global prediction of 101 km s$^{-1}$ from the tensor virial theorem.

## 4. SUMMARY

The main objective of this *Letter* is to demonstrate that the bulge mass may be higher than previously thought and that if the bulge is strongly barred, its total mass will depend sensitively on the amount of figure rotation (pattern speed) and its orientation to our line of sight.

For the parameters adopted here, the mass of the Galactic bulge is estimated to be $2.8 \times 10^{10}$ M$_\odot$, about 1.4 times higher than previous bar model estimates. The mass estimate is consistent with the mass which would be inferred from the observed optical depth to microlensing for stars in the bulge. The mass estimate may be slightly overestimated since I have ignored the contribution of the disk in the calculation. A similar calculation by the tensor virial theorem for the Kent (1992) model results in a mass estimate which is 20% higher than Kent's dynamical estimate.

The bulge mass derived here depends on factors which are not precisely known, but which are observationally constrained. The particular density distribution that I have used is only one of a number which fit the observed near infrared surface brightness distribution. The mass estimate is most sensitive to the pattern speed and angle between the line of sight and bar major axis. The adopted pattern speed is not directly observed through the motion of stars, but it results from a dynamical model which fits the observed gas kinematics well. If the pattern speed is significantly less than the value adopted here (81 km s$^{-1}$kpc$^{-1}$) or the angle between the line of sight and bar major axis is significantly larger than 20°, then the mass estimate would be similar to previous estimates for a barred bulge ($1.6 - 2.0 \times 10^{10}$ M$_\odot$).

The author is grateful to A. Gould for interesting and helpful discussions related to this work. The author also acknowledges the help of an anonymous referee whose suggestions have resulted in an improved paper. This work was supported by a National Science Foundation grant (AST − 9115236).

---